\documentclass[journal=jacsat,manuscript=article]{achemso}
\usepackage[version=3]{mhchem}

\author{Ming Li}
\affiliation{Key Laboratory of Quantum Information, University of Science and Technology of China, CAS, Hefei, 230026, People's Republic of China}
\alsoaffiliation{Synergetic Innovation Center of Quantum Information $\&$ Quantum
Physics, University of Science and Technology of China, Hefei, Anhui
230026, China}
\altaffiliation{Contributed equally to this work}
\author{Chang-Ling Zou}
\affiliation{Key Laboratory of Quantum Information, University of Science and Technology of China, CAS, Hefei, 230026, People's Republic of China}
\alsoaffiliation{Synergetic Innovation Center of Quantum Information $\&$ Quantum
Physics, University of Science and Technology of China, Hefei, Anhui
230026, China}
\altaffiliation{Contributed equally to this work}
\author{Xi-Feng Ren}
\affiliation{Key Laboratory of Quantum Information, University of Science and Technology of China, CAS, Hefei, 230026, People's Republic of China}
\alsoaffiliation{Synergetic Innovation Center of Quantum Information $\&$ Quantum
Physics, University of Science and Technology of China, Hefei, Anhui
230026, China}
\email{renxf@ustc.edu.cn}
\author{Xiao Xiong}
\affiliation{Key Laboratory of Quantum Information, University of Science and Technology of China, CAS, Hefei, 230026, People's Republic of China}
\alsoaffiliation{Synergetic Innovation Center of Quantum Information $\&$ Quantum
Physics, University of Science and Technology of China, Hefei, Anhui
230026, China}
\author{Yong-Jing Cai}
\affiliation{Key Laboratory of Quantum Information, University of Science and Technology of China, CAS, Hefei, 230026, People's Republic of China}
\alsoaffiliation{Synergetic Innovation Center of Quantum Information $\&$ Quantum
Physics, University of Science and Technology of China, Hefei, Anhui
230026, China}
\author{Guo-Ping Guo}
\affiliation{Key Laboratory of Quantum Information, University of Science and Technology of China, CAS, Hefei, 230026, People's Republic of China}
\alsoaffiliation{Synergetic Innovation Center of Quantum Information $\&$ Quantum
Physics, University of Science and Technology of China, Hefei, Anhui
230026, China}
\author{Li-Min Tong}
\affiliation{State Key Laboratory of Modern Optical Instrumentation, Department
of Optical Engineering, Zhejiang University, Hangzhou 310027, China}
\author{Guang-Can Guo}
\affiliation{Key Laboratory of Quantum Information, University of Science and Technology of China, CAS, Hefei, 230026, People's Republic of China}
\alsoaffiliation{Synergetic Innovation Center of Quantum Information $\&$ Quantum
Physics, University of Science and Technology of China, Hefei, Anhui
230026, China}

\title{Transmission of photonic quantum polarization entanglement in a nanoscale
hybrid plasmonic waveguide }

\makeatother

\begin{document}
\newpage{}
\begin{abstract}
Photonic quantum technologies have been extensively studied in quantum
information science, owing to the high-speed transmission and outstanding
low-noise properties of photons. However, applications based on photonic
entanglement are restricted due to the diffraction limit. In this
work, we demonstrate for the first time the maintaining of quantum
polarization entanglement in a nanoscale hybrid plasmonic waveguide
composed of a fiber taper and a silver nanowire. The transmitted state
throughout the waveguide has a fidelity of $0.932$ with the maximally
polarization entangled state $\Phi^{+}$. Furthermore, the Clauser,
Horne, Shimony, and Holt (CHSH) inequality test performed, resulting
in value of $2.495\pm0.147>2$, demonstrates the violation of the
hidden variable model. Since the plasmonic waveguide confines the
effective mode area to subwavelength scale, it can bridge nanophotonics
and quantum optics, and may be used as near-field quantum probe in
a quantum near-field micro/nano-scope, which can realize high spatial
resolution, ultra-sensitive, fiber-integrated, and plasmon-enhanced
detection.
\end{abstract}
\maketitle

\section{Keywords}

Surface plasmon polaritons, Quantum entanglement, Silver nanowire

\section{Introduction}

Quantum entanglement is one of the most extraordinary phenomena borne
out of quantum theory \cite{horodecki2009quantum}, and lies at the
heart of quantum information and future quantum technologies. From
a fundamental aspect, it is used to prove the Copenhagen interpretation
of quantum mechanics, against the famous EPR paradox \cite{clauser1969proposed,Bell}.
Quantum entanglement has powerful applications in information processing
and communications, as well as in the enhanced precision of measurement
\cite{giovannetti}. Quantum metrology \cite{giovannetti}, as an
important application of quantum entanglement, is the measurement
of physical parameters with enhanced resolution and sensitivity, enabled
by taking advantage of quantum theory, particularly by exploiting
quantum entanglement. For example, phase measurement with super-sensitivity
beyond the standard quantum limit (SQL) can be realized by using $N$-particles
entangled state $(N\geq2$) \cite{rarity1990two,kuzmich1998sub,nagata2007beating,xiang}.
This has many important applications, including gravitational wave
detection, measurements of distance and optical properties of materials,
and chemical and biological sensing. Very recently, entanglement-enhanced
microscopes, which give a signal-to-noise ratio better than that limited
by the SQL \cite{ono2013entanglement,Israel}, have been realized.
In these experiments, structures were measured using the far field
method, and the entangled photons were focused by an objective lens
onto the sample surface. Therefore, the spatial resolution of this
type of optical imaging system was fundamentally limited by the well-known
Abbe diffraction limit.

\begin{figure*}[htb]
\includegraphics[width=14cm]{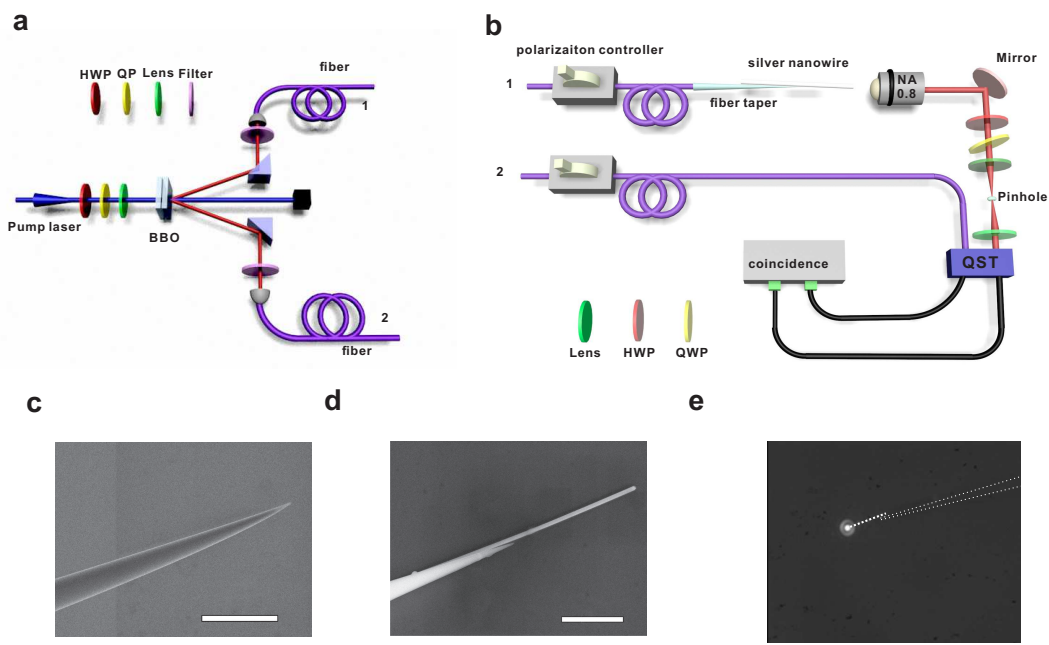} \caption{(color online) Experimental set-up. (a) The polarization entangled
photon source generated by a degenerate type-I non-collinear spontaneous
parametric down-conversion (SPDC) process. The produced photon pairs
(808 nm) are separated in free space by a $6^{\circ}$ angle based
on the phase-matching condition and directed to different optical
single-mode fibers. (b) One of the single mode fibers is connected
with a tapered fiber or further coupled with a silver nanowire, output
photons are collected by an objective lens and sent to the quantum
state tomography (QST) apparatus, and the other fiber is directly
connected to the QST. QP: Quartz plate; HWP: half-wave plate; QWP:
quarter-wave plate; BBO: $\mathrm{\beta-BaB_{2}O_{4}}$. (c) Scanning
electron microscope (SEM) image of a tapered fiber with a tip radius
of about $60$ nm. Scale bar: $5\ \mathrm{\mu m}$. (d) SEM image
of a fiber taper coupled with a $160$ nm radius silver nanowire.
Scale bar: $5\ \mathrm{\mu m}$. (e) CCD image of the hybrid waveguide.
Laser light is guided through the fiber taper, then coupled to the
silver nanowire and finally re-radiated from the nanowire end. }
\end{figure*}

To overcome this limitation, near field optics, such as a near-field
scanning optical microscope (NSOM), were developed. Typically, dielectric
tips (silica taper or nano-aperture) are used to probe nano-structures
beyond the diffraction limit. However, these near field probes suffer
from ultralow transmittance, generally only about $10^{-3}$. An alternative
method is to utilize surface plasmon polaritons (SPPs) \cite{barnes2003surface},
where the collective oscillations of free electrons are unlimited
by the diffraction. Not only can SPPs confine light beyond the diffraction
limit, but they can also enhance the light-matter interaction \cite{ozbay}.
Among various plasmonic structures, silver nanowires are a natural
choice for practical applications for several reasons: they are easy
to prepare, have regular and uniform geometry, and undergo relatively
low losses. Recently, great progress has been achieved in implementing
silver nanowire photonics for various applications \cite{xiong,guo2013nanowire},
such as waveguides, compact logic gates, single-photon sources, nanoscale
sensing and even single photon-level transistors.

Besides the studies on the classical applications of plasmonic structures,
researchers have started to investigate plasmonics in the quantum
regime. It has been experimentally proved that quantum polarization
entanglement \cite{altewischer2002plasmon}, energy-time entanglement
\cite{fasel2005energy}, and orbital angular momentum entanglement
\cite{ren2006} can be preserved in the photon-SPP-photon conversion
process through metal hole arrays. But the metallic film thickness
is less than optical diffraction limit (usually only about 100 nm),
which limits its application in plasmonic and quantum information
fields. More recently, quantum statistical property of single plasmon
was verified with metallic stripe waveguides\cite{qspnl} and bosonic
nature of it was also proved in on-chip Hong-Ou-Mandel experiments
\cite{nanotech,natphoton,prap,cai} based on plasmonic waveguides,
which show us the feasibility of achieving basic quantum logic gates
for linear optical quantum computation. However, the excitation of
SPPs is polarization-dependent, which may be an obstacle in the way
of transmitting polarization entanglement in plasmonic waveguide.
It is still unknown and remains as an experimental challenge whether
quantum polarization entanglement can be maintained at the nanoscale
plasmonic waveguides. Fortunately, a free-standing silver nanowire
may be a good candidate to answer this question, since it supports
not only the fundamental mode, but also two orthogonal higher-order
modes that can match the polarization of photons (see the Supporting
information), which outperforms the widely studied dielectric-loaded
SPP waveguide \cite{Kuma} and the metal strip waveguide \cite{Derigon},
for which SPPs can only be excited with transverse magnetic (TM) mode
light.

In this work, we studied the transmission of a photonic quantum-entangled
state through a nanoscale hybrid plasmonic waveguide, composed of
a silica tapered fiber and a silver nanowire. By performing quantum
state tomography \cite{qst} and a CHSH inequality test, preservation
of the quantum polarization entanglement in the waveguide is demonstrated
unambiguously, pushing the way forward to utilize quantum polarization
entanglement in plasmonic field. Importantly, this quantum entanglement-maintaining
nano-scale waveguide is fiber integrated, highly efficient, broadband,
robust and free to move. Therefore, it is a perfect candidate to be
used as near-field quantum probe for NSOMs and endoscopes \cite{Yang,Lu},
and may also be useful for quantum photonic integrated circuits \cite{obrien,cai,gate}.

\begin{figure}[htb]
\includegraphics[width=8cm]{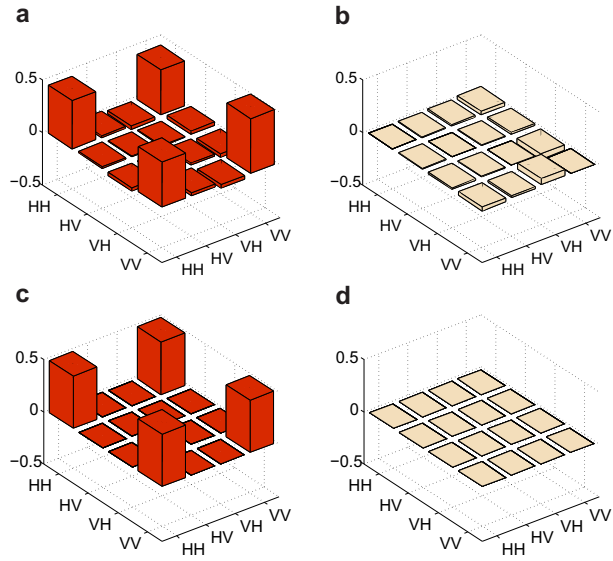} \caption{(color online) Density matrix of the output state from the fiber taper
and $\Phi^{+}$. (a) and (b) Real and imaginary parts of the density
matrix of the output state from the fiber taper. (c) and (d) Real
and imaginary parts of the density matrix of $\Phi^{+}$.}
\end{figure}

\section{Experiment and Results}

In our experiment, the polarization entangled two-photon state
\begin{equation}
\Phi^{+}=\frac{1}{\sqrt{2}}(|HH\rangle+|VV\rangle),
\end{equation}
was used to investigate the preservation of the quantum entanglement
in nanoscale waveguides. $\Phi^{+}$ is a maximally polarization entangled
Bell state, where $H(V)$ denotes horizontal (vertical) polarization.
The entangled photon pairs were generated via a type-I spontaneous
parametric down conversion (SPDC) process as depicted in Figure 1(a).
The generated quantum state was characterized by quantum state tomography
(Figure 1(b)). The concurrence \cite{Hill} of the measured density
matrix was $0.924$ (it is $1$ for a maximally entangled state and
$0$ for non-entangled state), and has a fidelity of $0.976$ with
the ideal state $\Phi^{+}$ (see the Supporting information).

To excite SPPs on the silver nanowire efficiently and make the nanowire
free-standing, we used a fiber taper \cite{Wu} to connect and lift
the silver nanowire. As shown in Figure 1(c), the taper has a very
smooth surface, and the diameter is adiabatically decreased from 125
$\mu m$ to about $120$ nm with a cone angle of about $3.5^{\circ}$.
Therefore, the photons in fiber can be adiabatically focused into
the nanoscale taper tip with negligible loss. Due to the perfect cylindrical
symmetry of the taper, the polarization of the photons should be preserved
during the focusing process. The maintaining of the quantum entanglement
through the taper was verified firstly by sending one of the two entangled
photons into the tapered fiber and the other through a single mode
fiber, as illustrated in Figure 1(b). The re-radiated photon from
the fiber taper was collected using an objective lens ($NA=0.8$),
and a confocal system composed of two lenses and one pinhole. The
pinhole was used to block the scattered background light and enhance
the signal to noise ratio. The quantum correlation between the two
photons was then investigated by quantum state tomography. The real
and imaginary parts of the measured density matrix are shown in Figure
2(a) and 2(b), and exhibit good agreement with the corresponding parts
of $\Phi^{+}$ (Figure 2(c) and 2(d)). Further calculations showed
that the quantum concurrence was 0.852 and the fidelity of the output
state with $\Phi^{+}$ was $0.958$.

Additionally, CHSH inequality tests were performed to examine the
non-local feature of the output photons. A hidden variable model requires
that
\begin{equation}
|S|=|E(\hat{A_{1}}\hat{,B_{1}})-E(\hat{A_{1}}\hat{,B_{2}})+E(\hat{A_{2}}\hat{,B_{1}})+E(\hat{A_{2}}\hat{,B_{2}})|\leq2,
\end{equation}
where $E(\hat{A_{i}}\hat{,B_{j}})$ ($i,j=1,2$) is the expectation
value of operator $\hat{A_{i}}\hat{B_{j}}$, and $\hat{A_{i}}$ and
$\hat{B_{j}}$ are the measurement operators on the two photons, respectively.
$\hat{A_{1}}$, $\hat{B_{1}}$, $\hat{A_{2}}$ and $\hat{B_{2}}$
correspond to $|H\rangle\langle H|$, $|A\rangle\langle A|$, $|V\rangle\langle V|$
and $|D\rangle\langle D|$, respectively, while $|A\rangle$ denotes
$\frac{1}{\sqrt{2}}(|H\rangle+|V\rangle)$, $|D\rangle$ denotes $\frac{1}{\sqrt{2}}(|H\rangle-|V\rangle)$
(see the Supporting information for detail). The maximal value of
$S$ in our experiment was $2.588\pm0.141$, which definitely violates
the hidden variable model. This further confirmed that the output
photons were still entangled after emerging from the taper.

A fiber taper is unique for the efficient coupling with other photonic
micro/nano-structures, because of the strong evanescent field around
the taper (see the Supporting information). Therefore, it is natural
to hybridize this dielectric waveguide with a SPP nano-structure.
Here, a silver nanowire was adopted. As shown in Figure 1(c) and 1(d),
the silver nanowire was adhered to the tip of the fiber taper, and
the coupling efficiency was high. In addition to the nano-scale photonic
confinement, there are several other benefits from such a hybrid dielectric-metal
nanotip: (1) Compactness. The highly efficient light coupling between
the silica taper and silver nanowire can be achieved within the wavelength
scale coupling region, avoiding harmful absorption loss in metal \cite{dong2009coupling-1,Guo}.
(2) In-line geometry. Photons from a single mode fiber can be efficiently
focused onto the tip of the silver nanowire. The transmission coupling
efficiency from the fiber to the silver nanowire was estimated to
be as high as $40\%$ in our experiment. (3) Robustness. The structure
is very robust and free to move; therefore, it has the potential to
be used in a quantum endoscope \cite{Yang,Lu}.

\begin{figure*}[htb]
\includegraphics[width=14cm]{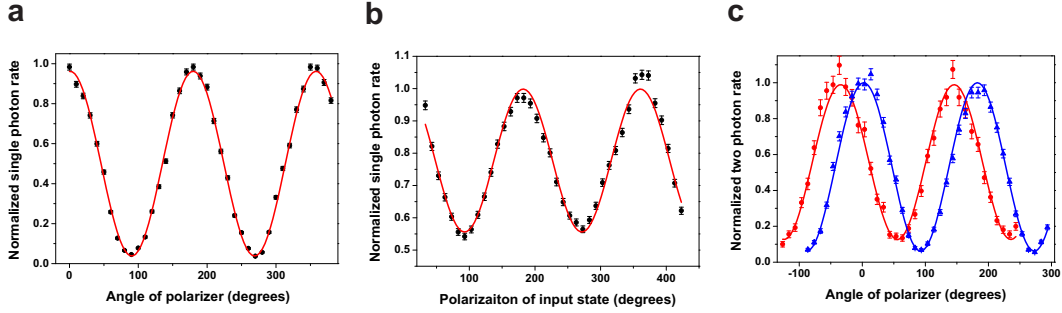} \caption{(color online) Single and biphoton fringes for the case of the silver
nanowire. The points are experimentally measured data and are fitted
with a sinusoidal function. (a) Single photon polarization property
of the fiber-nanowire structure. After injecting horizontally-polarized
photons, we measured the output counts while changing the orientation
of the polarizer in steps of $10^{\circ}$, where $0^{\circ}$ corresponds
to measurement of the $H$ state. (b) Transmission property of the
fiber-nanowire structure for different linearly polarized photons.
We measured the output counts while changing the polarization of the
input photons in steps of $10^{\circ}$, where $0^{\circ}$ denotes
the $H$ state. (c) Biphoton fringes corresponding to fourth-order
quantum interference. Blue and red dots are coincidence rates when
one photon was projected to a different polarization state, while
the other is projected to the $H$ and $\frac{1}{\sqrt{2}}(|H\rangle+|V\rangle$
states, respectively. Error bars are calculated from Poissonian statistics.}
\end{figure*}

Before testing the preservation of the quantum entanglement, we first
investigated the transmission of the hybrid tip using single photons.
The reason for doing so was that the coupling between the fiber taper
and the silver nanowire was a two-mode to three-mode process (see
the Supporting information), and the contact region of the silver
nanowire and the silica fiber tip was not cylindrically symmetric
(see Figure 1(d)), which influences the coupling between the dielectric
waveguide mode and the SPPs. First, $H$-polarized single photons
were sent into the hybrid tip, and the polarization of the transmitted
photons were analyzed by a half wave plate (HWP) and a polarizer.
The results are shown in Figure 3(a). The extinction ratio was measured
to be $25:1$, which indicates that the $H$-polarization was preserved
throughout the entire process (propagation of the photons in the silica
fiber, adiabatic focusing of the photons in the taper, conversion
between the photons and plasmons, propagation of the plasmons in the
silver nanowire, and scattering of the plasmons into free space photons
at the end of the nanowire). Secondly, the coupling efficiencies for
photons with different polarization were measured (Figure 3(b)). Due
to the asymmetric structure in the contact area, the coupling efficiency
changed with the polarization of the input photons and the ratio between
the $H$ and $V$ polarized photons was approximately 1.78, close
to the calculated result of $1.59$ (see the Supporting information).
The oscillations in the curve will be eliminated if the coupling efficiency
is identical for $H$ and $V$ polarized photons.

\begin{figure}[htb]
\includegraphics[width=8cm]{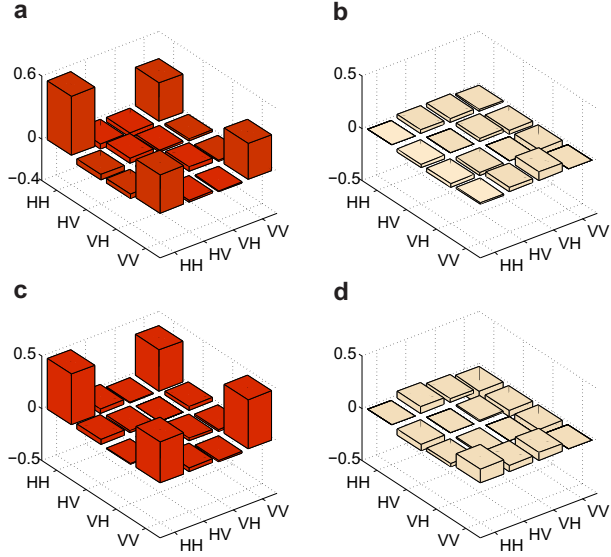} \caption{(color online) Density matrix of the output state for the case of
the silver nanowire. (a) and (b) Real and imaginary parts of the density
matrix of the original state from the silver nanowire. (c) and (d)
Real and imaginary parts of the density matrix of the output state
after an operation on the polarization of the pump laser to compensate
the unbalanced efficiencies of different polarizaitons. The output
state has a fidelity of 0.932 with the maximally entangled state $\Phi^{+}$.}
\end{figure}

Entangled photons were then used to test the hybrid tip, using an
experimental set-up similar to that used for the fiber taper. To show
the entanglement of the output photons intuitively, we measured the
coincidence rate by projecting one photon to a fixed state, while
scanning the projection state of the other. If the photons are entangled,
we can always find special projection states to make the coincidence
be $0$. For example, for the maximally entangled state $\Phi^{+}$,
when one photon is projected to $|D\rangle=\frac{1}{\sqrt{2}}(|H\rangle+|V\rangle$),
the other photon will be correspondingly collapsed to $|D\rangle$.
Therefore, the coincidence rate will become $0$ when we measure the
other photon using the $\frac{1}{\sqrt{2}}(|H\rangle-|V\rangle$ basis.
Similarly, if we project the first photon to state $|H\rangle$, the
coincidence rate will become $0$ with the other measured basis as
$|V\rangle$. The experimental biphoton fringes are shown in Figure
3(c), which can be treated as intuitive evidence of entanglement.
Furthermore, QST and CHSH inequality tests were also performed. The
density matrix is shown in Figure 4(a) and 4(b). It has a concurrence
of $0.700$ and a fidelity of 0.924 with the eigenstate $\widetilde{\Phi}=0.801|HH\rangle+0.594|VV\rangle$,
which was calculated using the density matrix diagonalization method.
Therefore, we can consider $\widetilde{\Phi}$ approximately to be
the output state. The change of state from the initial state to non-maximally
entangled state $\widetilde{\Phi}$ came from the unbalanced coupling
efficiency of the $H$ and $V$ photons, which was consistent with
the measured value $1.78$ for single photons. To compensate the unbalanced
coupling efficiencies, we can adjust the ratio between $|HH\rangle$
and $|VV\rangle$ of the input state by changing the polarization
of the pump laser. In this way, the output state could be tuned to
the maximally entangled state $\Phi^{+}$, with a fidelity of $0.932$
and a concurrence of $0.824$ (Figure 4(c) and 4(d)). The parameter
of the CHSH inequality was $2.495\pm0.147$, indicating violation
of the hidden variable theory.

\section{Discussion}

In plasmonic experiments, loss has long been considered to be the
major roadblock for practical applications. Here, we give a brief
discussion on whether loss would influence the entanglement. In our
experiment, three processes are responsible for the losses: the coupling
between the fiber taper and the silver nanowire, the intrinsic loss
of the plasmonic modes when propagating along the nanowire and non-unity
collection efficiency. Due to the symmetric shape of the silver nanowire,
the two higher-order modes of the structure used in our experiment
are degenerate, thus they experience the same loss in the propagation
process. The same as the absorption of the metal nanowire, the collection
process is also polarization independent, so both of the two processes
have no influence on the polarization entanglement. While as mentioned
above, the fiber-nanowire coupling process is polarization dependent,
because the coupling region is asymmetric for two different polarizations.
Thus the input polarization state will be changed to a non-maximally
entangled state. Fortunately, this problem can be settled by adjusting
the input entangled state or the collection equipment to ensure that
the output photons are still maximally entangled as shown in Figure
4(c) and 4(d).

In our experiment, the total transmittance from the photon source
to the detector was about $7.5\%$ for $H$-polarized photons. Taking
into account of the collection efficiency of the objective lens, the
efficiency of the confocal system and the free-space-to-fiber coupling
efficiency (about $40.1\%$, $70.5\%$ and $70.2\%$, respectively),
the efficiency from the fiber taper to the silver nanowire is estimated
to be about $40.3\%$ (see the Supporting information). By adjusting
the shape of the fiber taper and the coupling length, this efficiency
can be improved to even higher than $90\%$ \cite{oefibercouple}.
Compared to general near field probes, this high efficiency provides
us an opportunity to extract and study weak signals in the near field.
In addition, the effective mode area of the silver nanowire was only
about $0.38\lambda^{2}$, corresponding to a spot with diameter of
$0.35\lambda$ (see the Supporting information). This mode area can
be much smaller if we use the fundamental mode \cite{Tong}.

In summary, we have experimentally realized the transmission of quantum
polarization entanglement in a hybrid nanoscale plasmonic waveguide.
This hybrid device can confine the effective mode area to subwavelength
scale and can be applied as a quantum probe to realize high spatial
resolution and highly sensitive measurements, or to perform remote
excitation and remote sensing with the help of quantum entanglement
\cite{ren2008remote}, and the high efficiency of the device makes
these applications quite feasible. Our studies encourage further investigations
of the quantum effect in nanostructures through the coupling of these
waveguides, thus bridging free space quantum optical techniques and
nanophotonics. For example, we can envision a quantum nanoscope that
simultaneously beats the diffraction limit and SQL by using the techniques
exploited here.

\section{Supporting Information}

The Supporting Information contains the fabrication method of the
silver nanowire and the fiber taper, a detailed description of quantum
state tomography and CHSH test, the reconstructed density matrix of
the entangled photon source and the numerical simulation of the modes
and the transmission properties of our structure. This material is
available free of charge via the Internet at http://pubs.acs.org.

\begin{acknowledgement}

This work was funded by NBRP (grant nos. 2011CBA00200 and 2011CB921200),
the \textquotedbl{}Strategic Priority Research Program(B)\textquotedbl{}
of the Chinese Academy of Sciences(Grant No. XDB01030200), NNSF (grant
nos.11374289), the Fundamental Research Funds for the Central Universities
(grant no, WK2470000012) and NCET. We thank Y. F. Huang, B. H. Liu,
G. Y. Xiang, F. W. Sun, Z. B. Hou, W. Fang, H. K. Yu and H. Y. Zhang
for useful discussion.

\end{acknowledgement}

\section{Notes}

The authors declare no competing financial interests .

\end{document}